%% file: main.tex
\documentclass[sigconf]{acmart}
\input{defs.tex}
\AtBeginDocument{%
  }

\setcopyright{acmlicensed}
\copyrightyear{2025}
\acmYear{2025}
\acmDOI{XXXXXXX.XXXXXXX}
\acmConference[Conference acronym 'XX]{Make sure to enter the correct
  conference title from your rights confirmation email}{June 03--05,
  2025}{Woodstock, NY}
\acmISBN{978-1-4503-XXXX-X/2018/06}

\begin{document}

\title{Figaro on GPUs: Two Tables}

\author{Ðorđe Živanović}
\email{dorde@linscale.com}
\orcid{0000-0001-6440-0677}
\affiliation{%
  \institution{LinScale}
  \city{Edinburgh}
  \country{United Kingdom}
}

\begin{abstract}
  This paper introduces the implementation of the Figaro-GPU algorithm for computing a QR and SVD decomposition over a join matrix defined by the natural join over two tables on GPUs.
  Figaro-GPU's main novelty is a GPU implementation of the Figaro algorithm \cite{olteanu2022givens, vzivanovic2022linear,olteanu2024givens}: symbolical transformations combined with the GPU parallelized computations.
  This leads to the theoretical performance improvements proportional to the ratio of the join and input sizes.
  In experiments with the synthetic tables, for computing the upper triangular matrix and the right singular vectors matrix, Figaro-GPU outperforms in runtime  NVIDIA cuSolver library for the upper triangular matrix by a factor proportional to the gap between the join and input sizes, which varies from 5x-150x for NVIDIA 2070 and up to 160x for NVIDIA 4080 while using up to 1000x less memory than the GPU cuSolver.
  For computing singular values, Figaro-GPU outperforms in runtime NVIDIA cuSolver library from 2.8x-31x for NVIDIA 4080.
\end{abstract}

\maketitle

\section{Introduction}
This paper introduces Figaro-GPU that builds on the Figaro algorithm \cite{vzivanovic2022linear, olteanu2022givens, olteanu2024givens}. The Figaro algorithm computes the upper triangular matrix in the QR decomposition using Givens rotations by reducing the amount of computations for QR over a join matrix by applying structure-aware symbolical computations.
Similar holds for the SVD decomposition.

\textbf{Applications.} QR and SVD decompositions lie at the core of many linear algebra techniques, and their machine learning applications \cite{stewart1998matrix,stewart2001matrix, matrix2016comp} such as the matrix (pseudo) inverse and the least-squares needed in the closed-form solution of linear regression.
An SVD decomposition can be used for the principal component analysis of a matrix.

The matrix $\mat{R}$ in a QR decomposition constitutes a Cholesky decomposition of $\transpose{\mat{J}}\mat{J}$, which is used for training (non)linear regression models using gradient descent \cite{Schleich:F:2016}. If the matrix $\mat{J}$ is square, the product of diagonal entries in the matrix $\mat{R}$ equals: the determinant of $\mat{J}$, the product of the eigenvalues of the matrix $\mat{J}$ (ignoring the sign), and the product of the singular values of $\mat{J}$.

\textbf{Theory.}
Let $\mat{A}$ denote a matrix of dimensions $m \times n$: $\mat{A} \in \matspace{m}{n}$. Let $\mat{A}_{i, :}$ denote $i$-th row from the matrix $\mat{A}$. Let $\mat{0}_{m \times n}$ represents a zero matrix of dimensions $m \times n$.
For two matrices $\mat{A} \in \matspace{m_1}{n_1}$ and $\mat{B} \in \matspace{m_2}{n_2}$, a \textbf{Cartesian product} is a matrix $\mat{C} \in \matspace{m_1 m_2}{(n_1 + n_2)}:$ $$\mat{C} = \mat{A} \times \mat{B} = \begin{bmatrix}
  \mat{C}^{(1)} \\
  \mat{C}^{(2)} \\
  \vdots \\
  \mat{C}^{(m_1)} \\
\end{bmatrix} \text{ where } \mat{C}^{(i)} = \begin{bmatrix}
  \mat{A}_{i, :} & \mat{B}_{1, :}\\
  \vdots \\
  \mat{A}_{i, :} &  \mat{B}_{m_2, :}\\
\end{bmatrix}, i \in \setr{m_1}. $$
Let us introduce two definitions.
For a matrix $\mat{A} \in \matspace{m}{n}$, a \textbf{QR head} operator $\headQRName$ returns a matrix $\mat{B} \in \matspace{1}{n}$ and $\mat{B} = \headQR{\mat{A}}$ such that $$\mat{B} = \frac{1}{\sqrt{m}}\sum_{i=1}^{m}\vect{A}_{i, :}.$$
  For a matrix $\mat{A} \in \matspace{m}{n}$, a \textbf{QR tail} operator $\tailQRName$ returns a matrix $\mat{B} \in \matspace{(m-1)}{n}$ and $\mat{B} = \tailQR{\mat{A}}$ such that
    $$\vect{B}_{i, :} = \frac{1}{\sqrt{i+1}}\Bigg(\sqrt{i}\, \vect{A}_{i+1, :} - \frac{1}{\sqrt{i}} \sum_{k=1}^{i}\vect{A}_{k, :}\Bigg) , \text{ for } i \in \set{1, 2, \ldots, m-1}. $$

\textbf{Claim 1.} For matrices $\mat{A} \in \matspace{m_1}{n_1}, \mat{B} \in \matspace{m_2}{n_2}$, there is a product of sequence of Givens rotations $\mat{G} \in \matspace{m_1 m_2}{(n_1 + n_2)}$ such that:
$$ \mat{G} \cdot (\mat{A} \times \mat{B}) = \begin{bmatrix}
  \mat{A} \cdot \sqrt{m_2} & \headQR{\mat{B}} \\
  \mat{0}_{1\times n_1} & \sqrt{m_1} \cdot \tailQR{\mat{B}} \\
  \mat{0}_{(m_1 - 1) (m_2 - 1) \times n_1} & \mat{0}_{(m_1 - 1) (m_2 - 1) \times n_2}
\end{bmatrix}$$
For a QR decomposition, the standard approach computes the upper triangular matrix in $\bigO(m_1 \cdot m_2 \cdot (n_1 + n_2)^2)$.
If we apply \textbf{Claim 1}, and continue applying Givens rotations by zeroing out remaining entries,  entry by entry, we can compute the upper triangular matrix in the time complexity $\bigO((m_1 + m_2) \cdot (n_1 + n_2)^2)$ \footnote{We can obtain even faster runtime: $\bigO(m_1 \cdot n_1^2 + m_2 \cdot n_2^2)$ modulo other terms, but for the sake of the simplicity we sue abovementioned formula. }. This approach is a simplification of a general Figaro algorithm.

We paralleize the simplified Figaro algorithm as Figaro-GPU on a GPU in the following way. First, we store the input matrices $\mat{A}$ and $\mat{B}$ row-major and compute the respective columns in the resulting heads and tails using one to one mapping to Cuda threads. Then we concatenate the heads and tails using Cuda kernels, change the row-major to column-major using cublasXgeam and finish the upper triangular matrix using cusolverDnXgeqrf. We compute singular values by using cusolverDnXgesvd method on the upper triangular matrix.

\section{Experiments}
We evaluate the runtime performance of Figaro-GPU \footnote{\href{https://github.com/popina1994/LinScale-Cuda}{Figaro-GPU}} against NVIDIA cuSolver that use versions 12.6 and 12.8 using the cuSolver C++ API. NVIDIA cuSolver implements the Householder algorithm for the QR decomposition of dense matrices~\cite{Householder58b}.

We ran experiments on NVIDA 2070 8GiB Cuda 12.8 WSL2 and NVIDIA 4080 16GiB Cuda 12.6 Ubuntu 20.04.

The performance numbers are averages over 4 consecutive runs.
We do not consider the time to load the database into VRAM and assume that all relations and the join output are sorted by their join attributes.
We use the synthetic datasets of relations $\mat{S},\mat T \in \matspace{m}{n}$, whose join is the Cartesian product.
The data in each column follows a uniform distribution in the range $(0, 1)$.

\textbf{Computing \textbf{R}.}
We can see in Figure \ref{fig:experiment1-synthetic}, as we scale the number of columns, the speed-up increases since we paralelize computations column-wise and the memory access pattern is L2-cache friendlier for the Figaro-GPU.
Furthermore, as we scale the number of rows, the speed-up only increases since the ratio of the join output and join input only increases.
\input{experments/qr/qr.tex}

\textbf{Computing Singular Values.}
We can see in Figure \ref{fig:experiment2-synthetic} that if we compute the singular values that the similar reasoning as for computing the upper triangular matrix holds.
\input{experments/qr/svd.tex}
\section{Related Works}
\textbf{QR decomposition.}
There are three classical approaches to QR decomposition of a matrix $\mat A$: using Givens rotations \cite{Givens58} --- zeros entry by entry of the matrix A only affecting the two rows, while keeping the existing zero entries in the matrix $\mat{A}$ unmodified;  using Householder transformations ~\cite{Householder58b} ---zeros multiple entries in a column at once while affecting all other entries in the affected rows, \emph{Gram-Schmidt process}~\cite{Gram83,Schmidt07} computes the orthogonal columns of $\mat Q$ one at a time by subtracting iteratively from the $i$-th column of $\mat A$ all of its projections onto the  previously computed $i-1$ orthogonal columns of $\mat Q$.

\textbf{SVD decomposition.}
The approach to computing the SVD that is closest to ours proceeds as follows \cite[p.~285]{matrix2016comp}, \cite{lawson1995solving}. It computes the matrices $\mat R$ and $\mat Q$ in the QR decomposition of $\mat{A}$,
followed by the computation of the SVD of
$\mat{R} =
\mat U_R \mat \Sigma_R \mat \transpose{V_R}$.
Finally, it computes the orthogonal matrix $\mat U = \mat Q \mat U_R$. The SVD of $\mat A$ is then $\mat U \mat \Sigma_R \mat \transpose{V_R}$.
When only singular values are required, the squareroot-free or dqds method is the most popular~\cite{rutishauser1954quotienten,fernando1994accurate}.

\section{Conclusions}
In this paper, we presented Figaro-GPU, a novel algorithm for computing QR and SVD decomposition on GPUs.
We experimentally verified that Figaro-GPU outperforms in the runtime performance NVIDIA cuSolver for computing the upper triangular matrix for: 2070 and 4080 and for the singular values for 4080.

\newpage

\bibliographystyle{ACM-Reference-Format}
\bibliography{sample-base}

\appendix

\end{document}

%% file: defs.tex
\usepackage[utf8]{inputenc}   
\usepackage[T1]{fontenc}
\usepackage{textcomp}         
\usepackage{hyperref}
\usepackage{graphicx}
\usepackage{subcaption}

\newcommand{\mat}[1]{\mathbf{#1}}
\newcommand{\matspace}[2]{{\mathbb{R}}^{ #1 \times  #2}}
\newcommand{\headQRName}[0]{\mathcal{H}}
\newcommand{\headQR}[1]{\headQRName\left(#1 \right)}

\newcommand{\tailQRName}[0]{\mathcal{T}}
\newcommand{\tailQR}[1]{\tailQRName\left(#1 \right)}
\newcommand{\set}[1]{ \{#1 \} }
\newcommand{\setr}[1]{ \{1, 2, \ldots, #1 \} }
\newcommand{\vect}[1]{\mathbf{#1}}
\newcommand{\transpose}[1]{{#1} ^ \top}
\newcommand{\bigO}[0]{\mathcal{O}}

%% file: experments/qr/qr.tex
\begin{figure}[t]
    \begin{subfigure}{1.0\linewidth}

    \begin{center}
        \begin{tabular}{|l||r@{\hspace*{.75em}}r@{\hspace*{.75em}}r@{\hspace*{.75em}}r|}
            \hline
            & \multicolumn{4}{c|}{\#columns}\\
            \hline
            \#rows & $4$ & $8$ & $16$ & $32$\\
            \hline\hline
            $100$   & 0.18 &   0.23 &      0.33 &      0.55 \\
            $200$ &  1.71 &   1.81 &      1.85 &      2.08 \\
            $400$ &  1.71 &   1.75 &      2.89 &      2.11 \\
            $800$ &  1.68          &   1.74 &      1.98 &      2.14 \\
            $1600$ &  1.72 &   1.89 &      2.1 &      2.73 \\
            \hline
        \end{tabular}\hspace*{1em}%
        \begin{tabular}{|r@{\hspace*{.75em}}r@{\hspace*{.75em}}r@{\hspace*{.75em}}r|}
            \hline
            \multicolumn{4}{|c|}{\#columns}\\
            \hline
            $4$ & $8$ & $16$ & $32$\\
            \hline\hline
            10.5          &       9.4 &       8 & 6.8\\
            0.85 &            1.1 &       1.93 & 2.37 \\
            0.9 &           1.3 & 2.4 & 6.4           \\
            1.1 & 2.1 &  7           &   28.6        \\
            2.8 & 8.4 &    27.5            &  91.3          \\
            \hline
        \end{tabular}
    \end{center}

    \begin{center}
      \begin{tabular}{|l||r@{\hspace*{.75em}}r|}
          \hline
          & \multicolumn{2}{c|}{\#columns}\\
          \hline
          \#rows & $64$ & $128$  \\
          \hline\hline
          $100$   & 1.1 &   1.52      \\
          $200$ & 2.6 &   3.58        \\
          $400$ & 2.65 &   3.65        \\
          $800$ & 2.72 &   4.12        \\
          $1600$ & 3.82 &   6.43        \\
          \hline
      \end{tabular}\hspace*{1em}%
      \begin{tabular}{|r@{\hspace*{.75em}}r|}
          \hline
          \multicolumn{2}{|c|}{\#columns}\\
          \hline
          $64$ & $128$ \\
          \hline\hline
                  11.3 &     11.1         \\
                  6.3 &     9.2         \\
                  11 &    17.2                   \\
                  39.8 &   58.5           \\
                  119.2 &  158.4                 \\
          \hline
      \end{tabular}
  \end{center}
  \caption{Performance Runtimes for NVIDIA 4080}
\end{subfigure}

\begin{subfigure}{1.0\linewidth}

    \begin{center}
        \begin{tabular}{|l||r@{\hspace*{.75em}}r@{\hspace*{.75em}}r@{\hspace*{.75em}}r|}
            \hline
            & \multicolumn{4}{c|}{\#columns}\\
            \hline
            \#rows & $2$ & $4$ & $8$ & $16$\\
            \hline\hline
            $100$   & 0.50 &   0.38 &      0.39 &      0.55 \\
            $200$ &  0.31 &   0.43 &      0.50 &      0.64 \\
            $400$ &  0.44 &   0.6 &      0.73 &      0.94 \\
            $800$ &  0.77          &   1.02 &      1.25 &      1.28 \\
            \hline
        \end{tabular}\hspace*{1em}%
        \begin{tabular}{|r@{\hspace*{.75em}}r@{\hspace*{.75em}}r@{\hspace*{.75em}}r|}
            \hline
            \multicolumn{4}{|c|}{\#columns}\\
            \hline
            $2$ & $4$ & $8$ & $16$\\
            \hline\hline
            5.0          &       7.1 &       8.8 & 7.3 \\
            8.9 &            7.6 &       9.5 & 12.2 \\
            10.6 &           12.0 & 16.9 & 24.5           \\
            13.6 & 19.0 &  30.3           &   54.4       \\
            \hline
        \end{tabular}
    \end{center}

    \begin{center}
      \begin{tabular}{|l||r@{\hspace*{.75em}}r|}
          \hline
          & \multicolumn{2}{c|}{\#columns}\\
          \hline
          \#rows & $32$ & $64$  \\
          \hline\hline
          $100$   & 0.93 &   1.59      \\
          $200$ & 1.01 &   1.83        \\
          $400$ & 1.35 &   2.2        \\
          $800$ & 1.89 &   2.68        \\
          \hline
      \end{tabular}\hspace*{1em}%
      \begin{tabular}{|r@{\hspace*{.75em}}r|}
          \hline
          \multicolumn{2}{|c|}{\#columns}\\
          \hline
          $32$ & $64$ \\
          \hline\hline
                  6.4 &     11.9         \\
                  15.2 &     22.1         \\
                  37.3 &    50.6                  \\
                  90.9 &   147.2           \\
          \hline
      \end{tabular}
  \end{center}
  \caption{Performance Runtimes for NVIDIA 2070}
\end{subfigure}
  \caption{Runtime performance of Figaro-GPU and cuSolver for computing $\mat R$ in the QR decomposition of the Cartesian product of two matrices. The numbers of rows and columns are per matrix; for matrices of $100$ rows $2$ columns , cuSOlvers's input is a $10000\times 4$ matrix. Left: Runtime performance of Figaro-GPU (ms). Right: Speed-up of Figaro-GPU over cuSolver (rounded to one digit). An empty cell means that cuSolver runs out of memory. }
  \label{fig:experiment1-synthetic}
  \end{figure}

%% file: experments/qr/svd.tex
\begin{figure}[htb!]
    \begin{center}

        \begin{tabular}{|l||r@{\hspace*{.75em}}r@{\hspace*{.75em}}r@{\hspace*{.75em}}r|}
            \hline
            & \multicolumn{4}{c|}{\#columns}\\
            \hline
            \#rows & $4$ & $8$ & $16$ & $32$\\
            \hline\hline
            $100$   & 1.21 &   1.47 &      2.13 &      3.95 \\
            $200$ &  2.79 &   3.21 &      3.97 &      5.17 \\
            $400$ &  2.86 &   3.21 &      3.65 &      5.47 \\
            $800$ &  2.75          &   3.15 &      4.64 &      4.91 \\
            $1600$ &  3.51 &   3.83 &      7.65 &      8.15 \\
            \hline
        \end{tabular}\hspace*{1em}%
        \begin{tabular}{|r@{\hspace*{.75em}}r@{\hspace*{.75em}}r@{\hspace*{.75em}}r|}
            \hline
            \multicolumn{4}{|c|}{\#columns}\\
            \hline
            $4$ & $8$ & $16$ & $32$\\
            \hline\hline
            7.3          &       6.1 &       5.1 & 3.3\\
            3.0 &            2.9 &       2.9 & 2.8 \\
            3.0 &           3.0 & 3.3 & 4.0 \\
            3.2 & 3.5 &  4.6           &   14.0        \\
            3.2 & 6.0 &    8.6            &  31.4          \\
            \hline
        \end{tabular}
    \end{center}
  \caption{Runtime performance of Figaro-GPU and cuSolver for computing the singular values in the SVD decomposition of the Cartesian product of two matrices. The numbers of rows and columns are per matrix; for matrices of $100$ rows $2$ columns , cuSolvers's input is a $10000\times 4$ matrix. Left: Runtime performance of Figaro-GPU (ms). Right: Speed-up of Figaro-GPU over cuSolver (rounded to one digit). An empty cell means that cuSolver runs out of memory.
  All experiments run on NVIDIA 4080. }
  \label{fig:experiment2-synthetic}
\end{figure}